# Joint Detection and Classification of Communication and Radar Signals in Congested RF Environments Using YOLOv8

Xiwen Kang
Intelligent Fusion Technology, Inc.
20410 Century Blvd.
Germantown, MD 20874

Hua-mei Chen
Intelligent Fusion Technology, Inc.
20410 Century Blvd.
Germantown, MD 20874

Genshe Chen
Intelligent Fusion Technology, Inc.
20410 Century Blvd.
Germantown, MD 20874

Kuo-Chu Chang
Department of SEOR
George Mason University
4400 Univ. Dr.
Fairfax, VA 22030

Thomas M. Clemons
Department of SEOR
George Mason University
4400 Univ. Dr.
Fairfax, VA 22030

*Abstract*—In this paper, we present a comprehensive study on the application of YOLOv8, a state-of-the-art computer vision (CV) model, to the challenging problem of joint detection and classification of signals in a highly dynamic and congested radio frequency (RF) environment. Using our uniquely created synthetic RF datasets, we were able to explore three different scenarios with congested communication and radar signals. In the first study, we applied YOLOv8 to detect and classify multiple digital modulation signals coexisting within a highly congested and dynamic spectral environment with significant overlap in both frequency and time domains. The trained model was able to achieve an impressive mean average precision (mAP) of 0.888 at an intersection over union (IoU) threshold of 50%, signifying its robustness against spectral congestion. The second part of our research focuses on the detection and classification of multiple polyphase pulse radar signals, including Frank code and P1 through P4 codes. We were able to successfully train YOLOv8 to deliver a nearly perfect mAP50 score of 0.995 in a densely populated signal environment, further showcasing its capability in radar signal processing. In the last scenario, we demonstrated that the model can also be applied to the multi-target detection problem for continuous-wave radar. The synthetic datasets used in these experiments reflect a realistic mix of communication and radar signals with varying degrees of interference and congestion—a setup that has been overlooked by many past research efforts, which have primarily focused on ML-based classification of digital communication signal modulation schemes. Our study demonstrated the potential of advanced CV models in addressing spectrum sensing challenges in congested and dynamic RF environments involving both communication and radar signals. We hope our findings will spur further collaborative efforts to tackle the complexities of congested RF spectrum environments.

## I. INTRODUCTION

Recent developments in deep learning models have revolutionized the field of RF signal processing. In the broad area of spectrum sensing, deep learning models such as Convolutional Neural Networks (CNNs), Recurrent Neural Networks (RNNs), Long Short-Term Memory networks (LSTMs), and transformer networks have been adopted to extract key RF signal parameters, such as signal modulation scheme, signal bandwidth/symbol rate, and carrier frequency offset [1] –[5]. Of these parameters, automatic modulation classification (AMC) of digital communication signals receives significant attention from researchers. Previous research efforts on AMC can be broadly categorized into two approaches. One method involves applying neural networks directly to the in-phase and quadrature (I/Q) samples, ensuring that no information is lost due to signal processing and that all available information is used for neural network training [6]. However, this direct training approach often demands substantial memory space and computational resources due to the large size of I/Q samples [7]. Alternatively, the second approach aims to extract essential features from the I/Q samples using preprocessing techniques such as short-time Fourier transform (STFT) or Wavelet transform, which filter the input signal while preserving its key features [8][9]. Subsequently, the spectrogram of the I/Q signal is processed by image processing neural networks like CNNs.

Most prior studies on modulation classification have relied on synthetically generated datasets. For instance, in [10], the

RadioML 2016.10A synthetic dataset was developed using readily available GNU Radio modules, consisting of 11 modulations (8 digital and 3 analog) across varying signal-to-noise ratios (SNR). This dataset also incorporated several nonidealities such as sample rate offset, carrier frequency offset, multipath fading effects and AWGN. The RadioML dataset was later expanded to include samples of 24 digital and analog modulation types obtained via over the air measurement (RadioML 2018.01A) [11]. The Sig53 and Wideband-Sig53 (WBSig53) synthetic datasets introduced in [12] and [13] further expanded the total number of modulation schemes and signal classes to 53. These datasets also introduced additional hardware-related impairments such as I/Q imbalance and RF roll-off to better emulate real-world RF environment effects. As an example of a real-world measurement based dataset, [14] conducted an extensive over the air measurement campaign to create a dataset comprising cellular signals from multiple frequency bands recorded at different locations. The SPREAD dataset in [15] and WBR-DE dataset in [16] are some of the latest datasets that included communication and radar signals in congested environments.

Despite these recent efforts to diversify training datasets, current research predominantly relies on datasets tailored for general (and civilian) RF environments, typically focusing on static and isolated signal scenarios. These datasets fail to capture the complex and congested nature of the RF spectrum prevalent in modern electronic warfare settings, characterized by dynamic interactions and multiple overlapping signals. Firstly, synthetic datasets generated from past research efforts have been limited to digital communication signals, overlooking the diverse range of RF signals present in real-world electronic warfare (EW) environments. These environments encompass not only communication signals but also radar and navigation signals [17]. Secondly, previous datasets have lacked overlapping signals in both the time domain and frequency domain. In reality, the EW environment is often highly congested and contested [18]. In this paper, we attempt to address this gap by leveraging our synthetic RF signal generator capable of simulating a diverse array of signal types (including analog communication, digital communication, pulse radar, continuous-wave radar, and navigation signals) with varying characteristics such as carrier frequencies, bandwidths, modulations, SNR, time slots of transmission, and interference levels. Importantly, our generator also introduces the random occurrence of time and frequency domain overlapped signals. This capability adds a layer of complexity that better reflects modern electronic warfare operational settings, making our dataset not only unique but also more challenging compared to those typically used in current research.

In [13], the authors applied state of the art CV models such as YOLOv5 (You Only Look Once) and DETR (Detection Transformer) to analyze the digital communication signals from their WBSig53 datasets. YOLOv8 developed by Ultralytics, represents the newest advancement in CV model, succeeding YOLOv5 [19]. According to Ultralytics, the latest YOLOv8 model incorporates various architectural enhancements and improvements in developer experience compared to YOLOv5.

In this paper, we will leverage YOLOv8 to tackle the challenging problem of jointly detecting and classifying overlapping communication and radar signals, a topic that has received limited attention in prior research efforts. Our contributions are threefold: Firstly, we have demonstrated that state-of-the-art CV models like YOLOv8 can be trained to offer a robust solution for simultaneous detection and classification of signals in highly dynamic and congested spectrum environments. This includes various types of digital modulation signals overlapping with each other in both frequency and time domains. Our trained YOLOv8 model was able to achieve a mean average precision (mAP) of 0.888 at 50% IoU.

Secondly, we applied the model to the detection and classification of polyphase pulse radar signals. Our model successfully located and classified multiple types of polyphase pulse radar signal, including the Frank code and P1 through P4 codes. We achieved a nearly perfect mAP score of 0.995 at a 50% IoU.

Thirdly, we were able to apply our model to address the multi-target detection problem for continuous-wave radar signals. In addition to these three results, our work produced sophisticated synthetic RF datasets containing a diverse array of signal types, simulating highly congested and dynamic spectrum environment. These datasets not only supported our experiments but also serve as valuable resources for future collaborative research efforts aimed at studying congested RF spectrum environments.

The paper is organized as follows: Section II introduces our synthetic RF signal generator, capable of simulating a dynamic, congested, and contested RF spectrum environment with multiple signal types. Section III details the experimental setup and the generation of our datasets. Section IV evaluates the performance and results of the YOLOv8-based approach. Finally in section V, we present our conclusions and explore potential directions for future research.

## II. SIGNAL SIMULATOR CAPABILITIES

### A. Capabilities and Input Parameters

The current version of the simulator, developed using MATLAB communication toolboxes and additional proprietary functions we implemented, supports the generation of 26 different modulations and signal types used in communication, radar, and navigation applications. The simulator is highly customizable, allowing users to set various transmitter signal parameters and configure the channel condition and propagation environment. In addition, the simulator offers a dynamic mode where key signal parameters, such as bandwidth/symbol rate, carrier frequency, signal power, and transmission duration, can be continuously changing to emulate a dynamic spectrum environment.

### B. Example Output

Once all transmitter nodes and channel conditions are configured, the simulator will execute to compute the received signal and visualize the results in both time and frequency domains. As an example, Fig. 1, shows the time domain waveform and frequency domain spectrogram of six different communication and radar signals computed using the short-time Fourier Transform (STFT). For easy interpretation of the time-domain waveform, all signals here have their carrier frequencies set to zero (i.e., at the baseband).

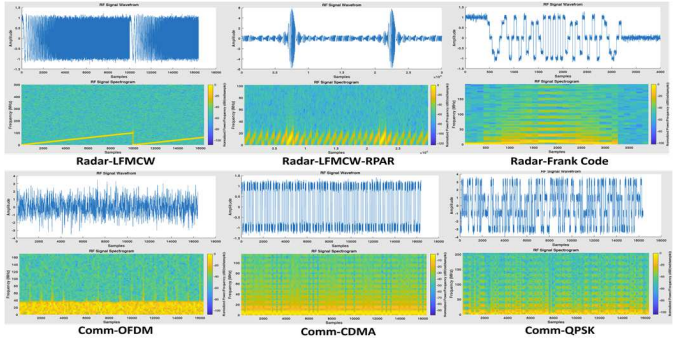

Fig. 1. Example output of the RF simulator (showing the baseband I-component only).

In addition to showing the signal received from each transmitter node individually, our simulator can compute and visualize the aggregate received signal from all transmitters based on their relative power. Fig. 2 illustrates a scenario with seven communication and radar transmitters coexisting within the 500 MHz band. To better visualize the congested spectrum, Fig. 3 presents each signal instance enclosed by a bounding box. The height of the box corresponds to the signal bandwidth/symbol rate, while the width represents the duration of the transmission. In the annotated spectrogram, multiple instances of signals overlapping in both time and frequency domains are evident. This complexity poses challenges for signal detection and classification tasks.

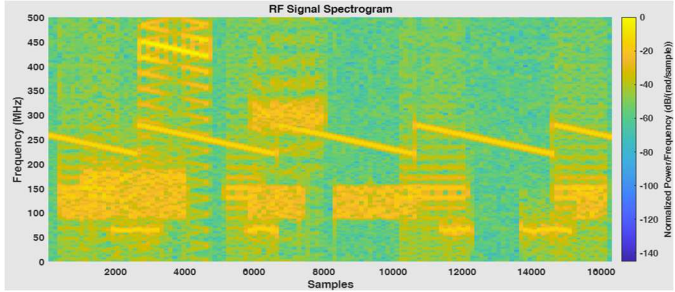

Fig. 2. Example of a congested RF environment.

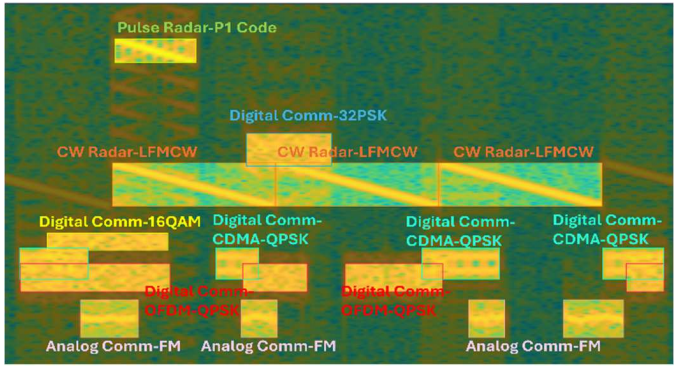

Fig. 3. Annotated spectrogram with bounding boxes showing different signal types coexisting with each other.

## III. EXPERIMENT SETUP

In this section, we will present the simulation setup and dataset generation process for each scenario under study.

### A. Congested spectrum with digital modulation signals

#### 1) Simulation Setup

Like many of the previous research efforts, our first scenario focuses purely on digital communication signals and aims to simulate a dynamic spectrum environment with eight different modulation types. To simulate a dynamic spectrum and learn a generalized model, we allow parameters such as SNR, bandwidth/symbol rate, carrier frequency and transmission duration to vary within certain intervals. The exact range for each parameter is summarized in TABLE I. below.

TABLE I. SIMULATION SETUP FOR CONGESTED DIGITAL MODULATION SIGNALS

| | |
|---|---|
| Modulation Schemes | QPSK, 8PSK, 16PSK, 32PSK, 16QAM, 32QAM, CDMA-QPSK, OFDM-QPSK |
| Number of Samples (per Spectrogram) | 4096 samples/timeslot * 4 timeslots = 16384 samples |
| Number of Spectrograms Generated | 4000 |
| Carrier Frequency (MHz) | 60 MHz < fc < 440 MHz |
| Single-sided Bandwidth/Symbol Rate (MHz) | 20 MHz < BW < 60 MHz |
| Transmission Duration (% of timeslot duration) | 20% < Dt < 100% |
| SNR (dB) | 0 dB < SNR < 25 dB |

Each spectrogram used for training includes 16,384 samples of the aggregate received signal, and we will demonstrate later that using only 4,000 of such spectrograms (divided into training, validation and testing sets) could yield very good detection and classification performance. The signal SNR is allowed to vary between 0 dB to 25 dB in order to learn a generalized model that works for both high SNR and low SNR scenarios. For the same reason, the carrier frequency is allowed to vary as much as the simulator sampling bandwidth allowed (i.e., within the 500 MHz range). The 60 MHz signal bandwidth upper bound is set so that the spectrum does not get unreasonably crowded. Additionally, we have set 20 MHz as the minimum signal bandwidth and 20% as the minimum transmission duration duty cycle within a transmission timeslot. This is to ensure that the size of the "image" is not unreasonably small for the CV model. Despite these posed constraints, the CV model could still apply to signals with narrower bandwidth or shorter duration if we reduce the observation window size in both time and frequency domains (i.e., the number of samples in each spectrogram and bandwidth of interest). Fig. 4 shows an example of the synthesized spectrogram comprising eight overlapping digital modulation signals.

#### 2) Dataset Generation

Since the ground truth carrier frequency, bandwidth and transmission duration of each signal is known during the synthesis process, we could easily create the corresponding bounding box in XYWH format with normalized coordinates given by:

$$\text{Horizontal position of the center: } X = \frac{\frac{t_{start}+t_{end}}{2}}{16384} \quad (1)$$

$$\text{Vertical position of the center: } Y = \frac{f_c}{500\ MHz} \quad (2)$$

$$\text{Image Width: } W = \frac{t_{end}-t_{start}}{16384} \quad (3)$$

$$\text{Image Height: } H = \frac{BW}{500\ MHz} \quad (4)$$

Once the bounding box labels are created, the 4,000 spectrograms are divided into training, validation and test sets, each containing 2,800, 800, and 400 images, respectively.

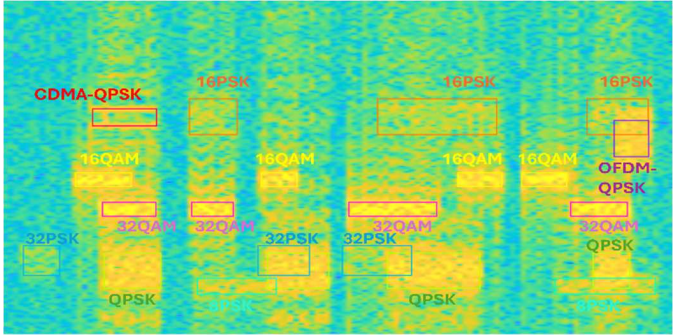

Fig. 4. Example of the generated congested spectrogram with 8 different modulation schemes and overlapping signals.

### B. Congested spectrum with pulse radar signals

#### 1) Simulation Setup

The simulation setup and input parameter range for the pulse radar scenario is the same as the digital communication scenario except that the radar signal is always ON (i.e., there is no medium access control or transmission timeslot). Our goal is to detect and classify five different polyphase pulse radar signals including Frank code and P1 through P4 codes. Fig. 5 shows an example of the synthesized spectrogram comprising five overlapping pulse radar signals.

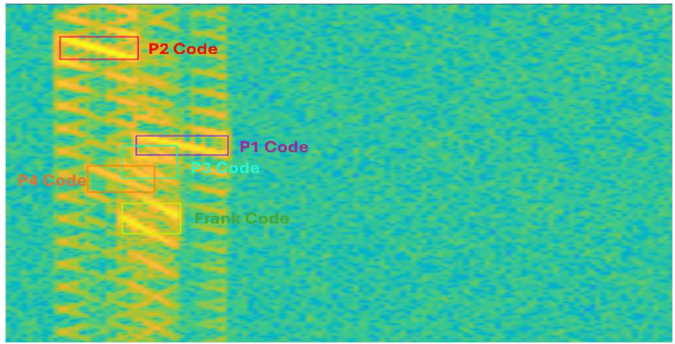

Fig. 5. Example of the congested spetrogram with 5 different polyphase pulse radar waveforms.

#### 2) Dataset Generation

The dataset generation and labeling process is the same as the digital communication scenario except that the signal transmission duration now becomes the pulse duration.

### C. Multi-target detection for continuous-wave radar

The YOLOv8 model could also be applied to the target detection problem for continuous-wave radar. Using our simulator, we are able to create a scenario where multiple overlapping echoes are received for a single LFMCW radar transmitter. Our goal is to train a model to automatically detect and locate all the echoes reflected from multiple targets. Fig. 6 shows an example of the spectrogram containing echoes from three different targets with various power levels.

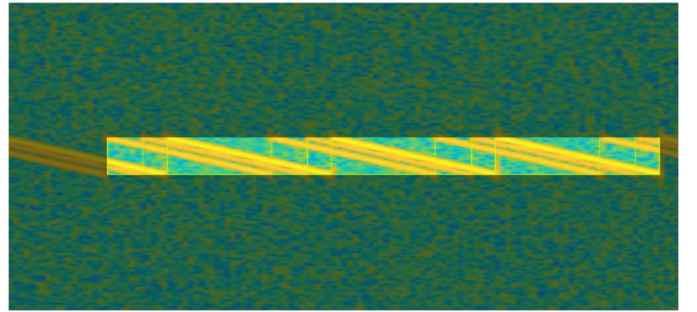

Fig. 6. Example of the spectrogram containing echoes (with varying power level) reflected from 3 different targets.

## IV. PERFORMANCE EVALUATION

For each of the three scenarios, we fine-tuned the Ultralytics YOLOv8s model on a single NVIDIA Tesla T4 GPU, utilizing a training set of 2,800 spectrogram images. The model was fine-tuned for 50 epochs, employing an AdamW optimizer with a batch size of 16 and a learning rate of 8.33e-4.

### A. Congested spectrum with digital modulation signals

With only 2,800 training spectrograms, we were able to fine-tune a generalized model (i.e., generalized SNR, symbol rate, carrier frequency, transmission time and duration) with 0.888 mAP50 score and 0.753 mAP50-95 score after 2.4 hours of training. Fig. 7 shows the normalized confusion matrix.

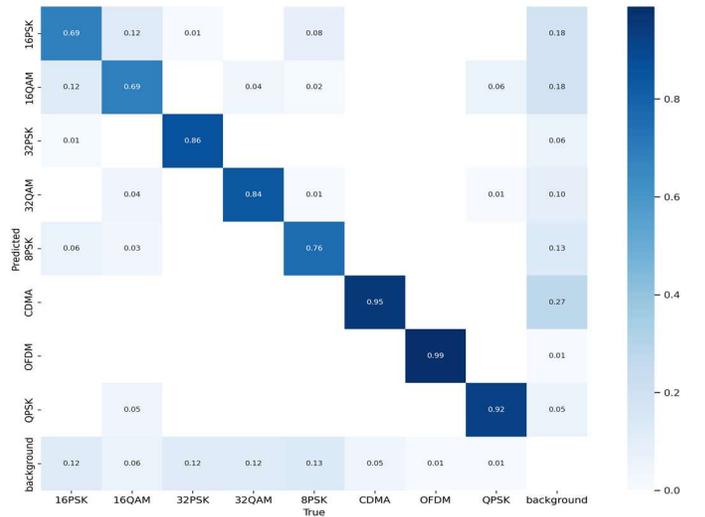

Fig. 7. Normalized confusion matrix for digital modulation detection and classification in a congested environment

The CV model had exceptional performance with CDMA and OFDM waveforms because their spectrogram fingerprints are more unique compared with the other six classes of signals being studied. It can also do a decent job in distinguishing signals from the same modulation family (e.g., M-PSK).

After completing training, the model was deployed to conduct inference on the test datasets. Example prediction results are depicted in Fig. 8 and Fig. 9. As observed, YOLOv8 can accurately detect and locate each signal with high confidence. Furthermore, the model demonstrates robust performance in scenarios involving time and frequency overlapped signals.

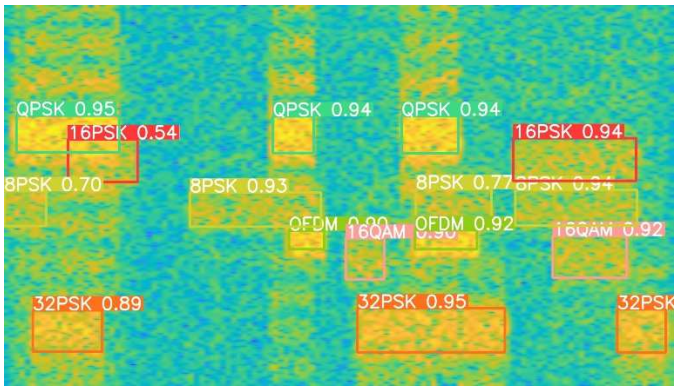

Fig. 8. Example detection and classification result for congested digital modulation signals

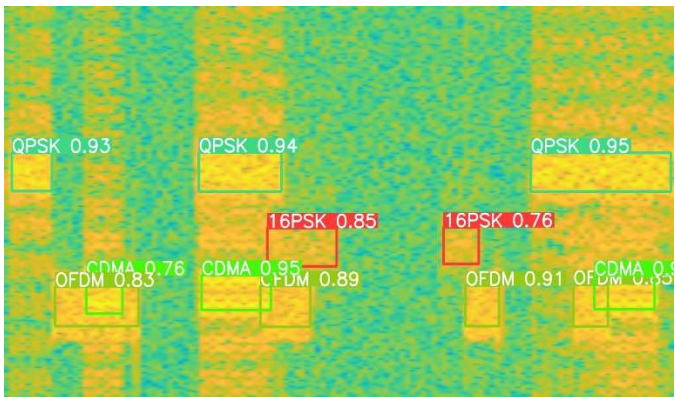

Fig. 9. Example detection and classification result for congested digital modulation signals

### B. Congested spectrum with pulse radar signals

For the pulse radar signals, we were able to fine-tune a generalized model using 2,800 spectrograms and obtain a nearly perfect mAP50 score of 0.995 and a mAP50-95 score of 0.968. Fig. 10 shows the normalized confusion matrix. The overall accuracy is much better than the digital modulation scenario because we have less number of signal classes which lead to less crowded spectrograms.

Fig. 11 shows an example detection and classification result by the trained model. Remarkably, even in a congested environment, the model remained robust and was able to locate and classify the pulse radar signals with high confidence, despite the overlapped spectrogram.

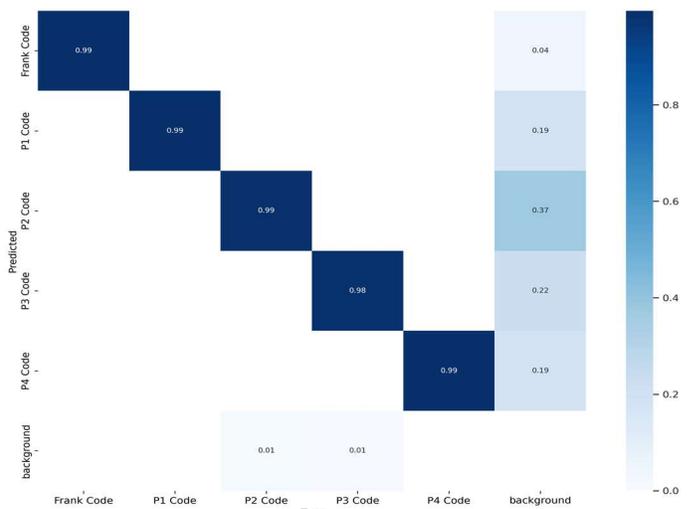

Fig. 10. Normalized confusion matrix for polyphase radar pulse signals detection and classification in a congested environment

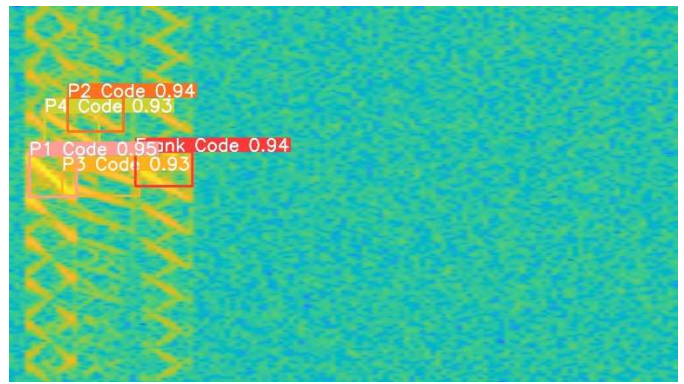

Fig. 11. Example detection and classification result for congested pulse radar signals

### C. Multi-target detection for continuous-wave radar

As mentioned earlier, we also explored the potential use of the YOLOv8 model to the problem of multi-target detection for continuous-wave radar applications.

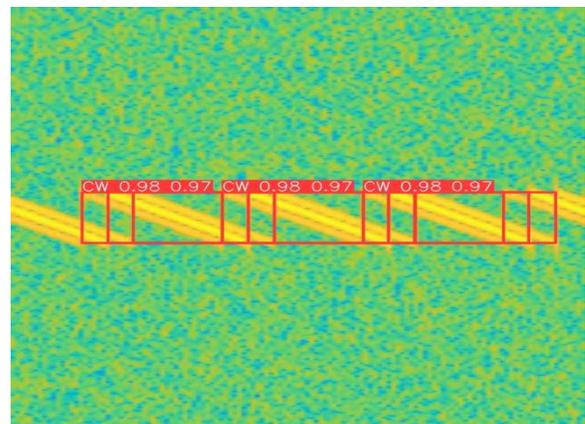

Fig. 12. Example multi-target detection result for continuous-wave radar

In continuous-wave radar, echoes reflected off multiple targets produce multiple chirp signals with varying power, overlapping in time and frequency domains within the same spectrogram. As illustrated in Fig. 12, the YOLOv8 model was successfully trained to automatically identify and locate all reflected chirp signals with high accuracy. This capability serves as the foundation for subsequent processing and separation of primary and secondary targets.

## V. Conclusion and Future Work

In this work, we demonstrated the capabilities of state-of-the-art (SoA) CV models such as YOLOv8 in handling dynamic and congested spectrum environments commonly encountered in modern electronic warfare scenarios. With a mean average precision (mAP50) of 0.888 in classifying eight digital modulation schemes and an impressive mAP50 of 0.995 for pulse radar signals, both within highly dynamic and congested spectrum environments, our research confirms that YOLOv8 can robustly manage the challenges posed by overlapping signals in both time and frequency domains.

This successful application of YOLOv8 to the signal detection and classification problem in congested spectrum has deeper implications beyond spectrum sensing. Firstly, it highlights the substantial potential of CV models as pre-processors for data-heavy, time-sensitive EW platforms. By accurately pinpointing signals' locations in time and frequency, selective filtering (followed by downconversion to baseband) and time-domain truncation can be applied to pre-process the I/Q or spectrogram input data. This approach conserves memory space and computational resources while retaining essential information for subsequent analysis. Secondly, the CV model has demonstrated its capability to extract essential features and metadata, such as carrier frequency, bandwidth/symbol rate, and modulation scheme, from the input samples or spectrograms. This ability enables subsequent processing by AI/ML-based pattern-of-life analysis platforms, enhancing battlefield command and control capabilities and supporting commanders in their decision making processes.

Regarding future research directions, we identify two potential areas for further studies. Firstly, we intend to implement a hardware prototype to validate the effectiveness and reliability of YOLOv8 and other state-of-the-art CV models in processing samples collected in real-world spectrum settings. Secondly, we recognize the potential to improve the model's overall detection and classification performance through advanced data augmentation techniques. In a separate paper, we will discuss how the choices of STFT parameters, such as the window length, window type, FFT size and overlap factor, could affect the overall accuracy and robustness of the computer vision-based RF signal detection and classification methods.

## VI. Software and Data

The datasets (spectrograms and bounding box labels) used for training and testing in each of the three scenarios are available for download at: https://github.com/xwkang2019/CongestedSpectrum.